# Electronic structure of barium titanate : an abinitio DFT study

**Hong-Jian Feng, Fa-Min Liu**

Department of Physics, School of Sciences, Beijing University of Aeronautics & Astronautics, Beijing 100083, P. R. China

**Abstract:**

First principle calculations were performed to study the ground state electronic properties of Barium titanate within the density functional theory (DFT). In our DFT computations, we used Vosko-Wilk-Nusair correlation energy functional and generalized gradient approximation (GGA) exchange and correlation energy functional as suggested by Perdew and Wang (PWGGA). The band structure, total density of states (DOS) and partial DOS have been systematically conducted to investigate the electronic configuration of this prototype ferroelectric perovskits compound. The band gap was 1.92 eV within our approach, and the quasi-flat band at -17 eV and -10 eV were attributed to the O 2s and Ba 5p states respectively, which was in good agreement with the corresponding total DOS and partial DOS. From the DOS investigation, it can be seen that the Ti $e_g$ state intended to interact with the oxygen octahedral orbitals to form the p-d hybridization. Moreover the strong p-d overlap and bonding can be observed in the electronic density redistribution along the different crystalline planes with respect to the corresponding space group, and the electronic isodense have been shown along the (001), (100), (110) and (111) crystal planes. From these electronic density maps, the strong bonding between Ti and O atoms can even be observed in the (111) crystalline plane.

## 1. Introduction:

Barium titanate (BTO) is a classical example of ferroelectric material which was widely used in electronic devices as high permittivity capacitors, infrared detectors or transducers due to its particular characteristics [1-6]. And it is of the perovskites $ABO_3$ type materials which have been intensive investigated for at least half a century. Much experimental works have been conducted to analyze the properties of the perovskites type ferroelectric materials in order to push forward the applicable progress. Meanwhile numerous theoretical works have been employed to discover the driven mechanism of the ferroelectric properties. In recent years, Magnetoelectric multiferroics have attracted tremendous interests due to their magnetoelectric properties, originating from the coupling between ferroelectric and ferromagnetic order parameter. [7-11] In order to find the intrinsic mechanism of these attractive properties, more theoretical calculations should be done to investigate the atomic and electronic structure of these series of materials. First principles has been recently used to describe the electronic structure of materials, and one of the first theoretical investigations in BTO ferroelectric transition has been done by Cohen and Krakouer in the early 1990s using the first principles calculations based on density functional theory (DFT) with local density approximation (LDA) method [12-13]. Nowadays, it is well-known that the DFT-LDA calculations could underestimate the band gap of experimental results by a factor of 2 [14]. On the other hand, the pure HF calculations always overestimate the experimental value. Therefore, these two methods should be mixed to achieve the improved approach to calculate the precise results.

Although the different calculations have been performed to specify the electronic structure of $ABO_3$ perovskites materials, unfortunately most of the lattice constant used in the previous work were acquired from the theoretical results of published papers. In this paper the lattice constant used have been mainly acquired from our X-Ray Diffraction results. Although most work have been conducted in the study of perovskites type barium titanate, fewer papers could be found to systematically describe the electronic configuration of this sort of material using the band structure, total DOS and partial DOS. Moreover more calculations should be done to describe the electronic density distribution along different crystalline planes in terms of the corresponding space group. In this paper we mainly focused on the calculations of the band structure, density of states (DOS) and the charge density along different crystalline planes within the DFT exchange and generalized gradient approximation (GGA) correlation functional to provide comprehensive understanding on the forthcoming experiments.

## 2. Computational details:

The calculation was separated into three steps. Firstly, the lattice constant optimization was performed with respect to the lowest energy. The initial lattice constant (a=3.9904Å, c=4.0689 Å) were provided, which have been derived from the average value calculated from our experimental samples, BaTiO3 powder and those doped with different amount of Fe, which can give the perfect average value. The X-ray diffraction pattern was shown in Fig.1.

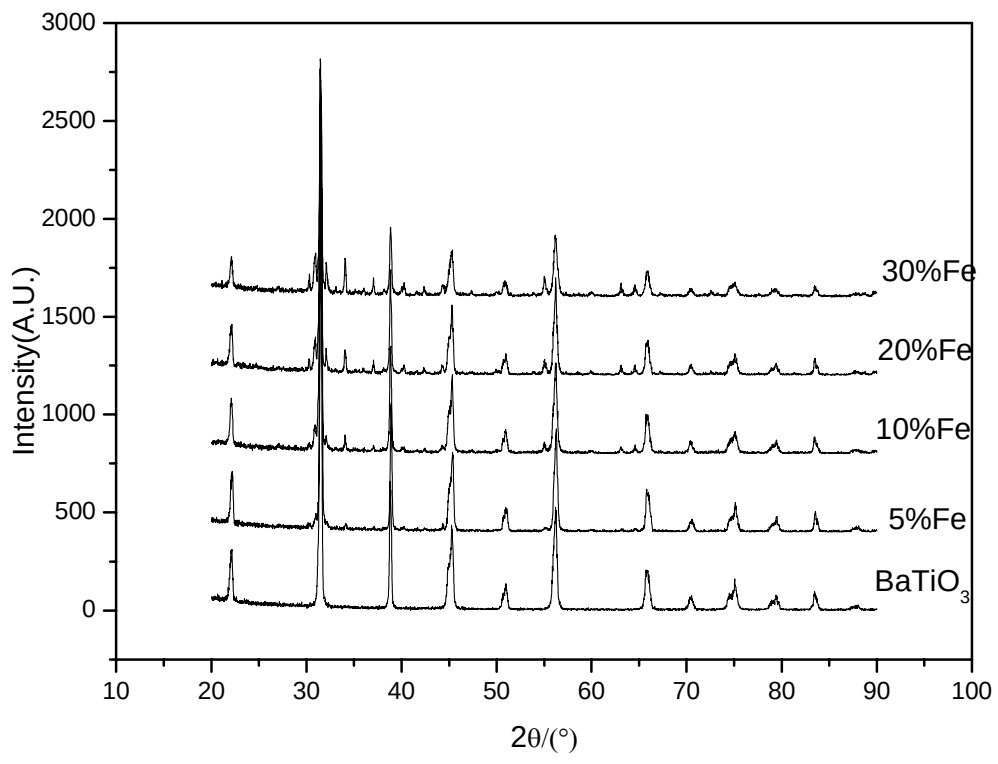

Fig. 1 X-ray diffraction patterns of different Fe doping BaTiO3

Secondly, we conducted the self-consistent calculation, and tetragonal phase is known experimentally to be stable at room temperature, and the corresponding space group was p4mm. Therefore, we constructed the tetragonal structure for $BaTiO_3$, and it is shown in Fig.2. During relaxation the cell shape and volume was fixed. In the last step, we performed the non-self-consistent calculation. In our DFT computations, we used Vosko-Wilk-Nusair correlation energy functional and generalized gradient approximation (GGA) exchange and correlation functional as suggested by Perdew and Wang (PWGGA). The reciprocal space integration was performed by sampling the Brillouin Zone with the 11×11×11 Monkorst-Pack net. The high symmetry points in the reciprocal space were selected to compute the band structure of barium titanate. Plane-wave functions were used as basis sets, and a plane-wave cutoff energy of 500 eV was employed throughout. It has shown that the results are well converged at this cutoff. Spin–orbit interaction was excluded in our calculations due to its weak influences in 3d elements. Ba 5s, Ba 5p, Ti 3s, Ti 3p, Ti3d, O2s and O2p electrons have been treated as valence state.

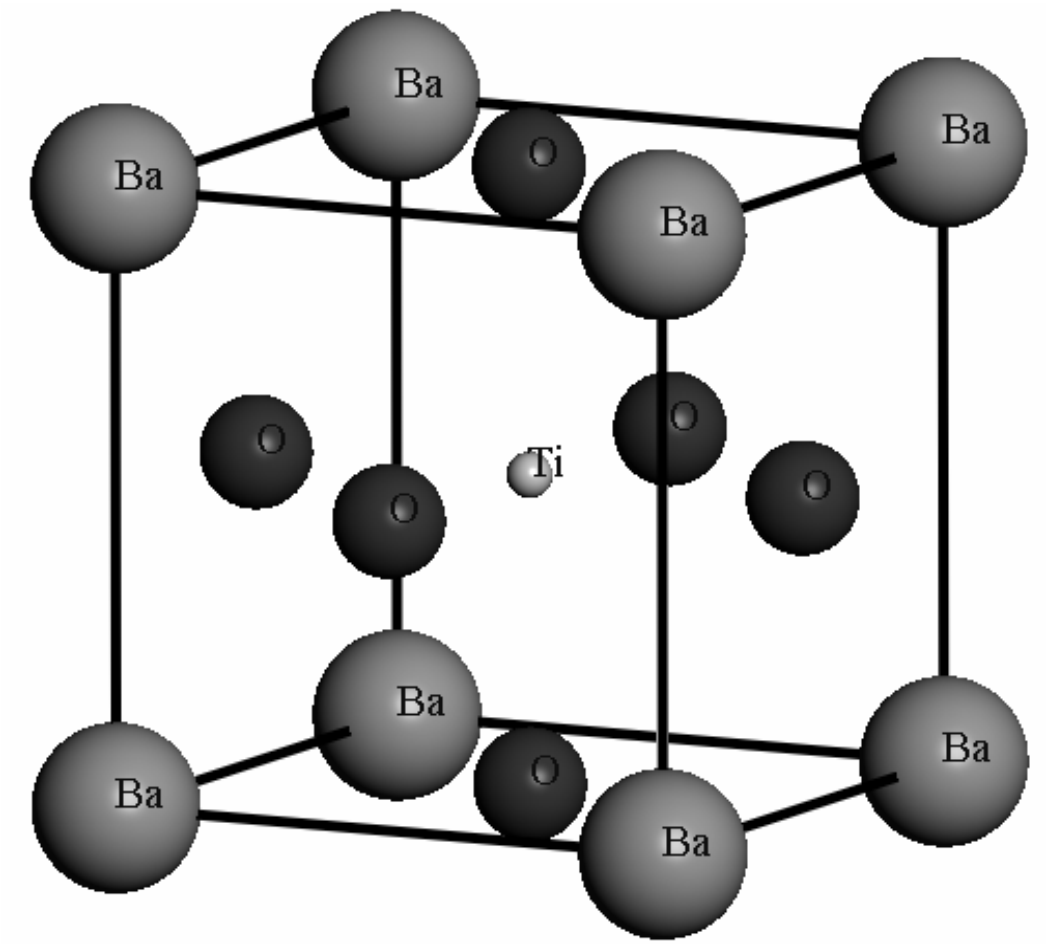

Fig. 2 The diagram of tetragonal structural $BaTiO_3$

## 3. Results and discussion

The calculated band structure of $BaTiO_3$ was shown in Fig. 3. The Band structure was separated by four portions in the valence band. The narrow band positioned at -17eV was derived from O2s states, and the one positioned at -10eV was originated from the Ba5p states. There is a manifold in the valence band near Fermi level which was attributed to the O2p states, moreover the conduction band near Fermi lever have a strong Ti 3d characteristic. These results could be confirmed by the total density of states (DOS) in Fig.4. The sharp peak at -17eV and -10eV denoted the corresponding two narrow bands. However, there was a sharp difference compared with the previous computed results, which an excess band positioned at -32eV was detected in our calculation, and the corresponding states can also be observed in the total DOS. This novel phenomenon might be caused by the O 2s states. The band gap was 1.92 eV within our approach, and this was in good agreement with the fact that the band gap was underestimated by a factor of 2 within the DFT-LDA method, and we improved the calculated value by 10% compared with the previous one [14].

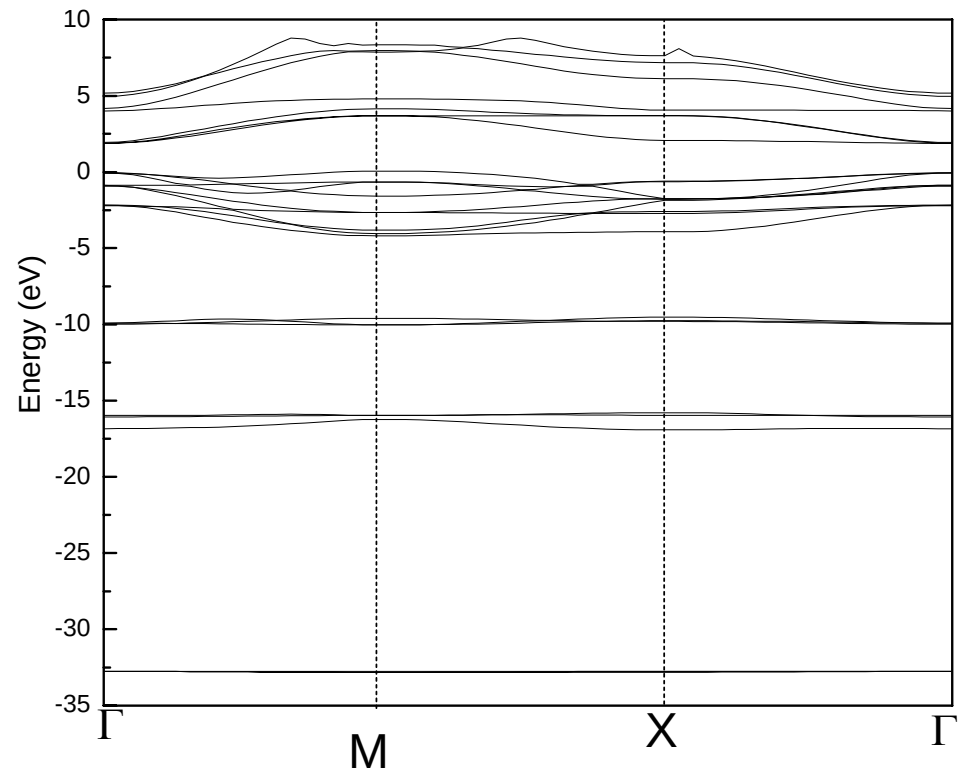

Fig. 3 The band structure calculated for $BaTiO_3$

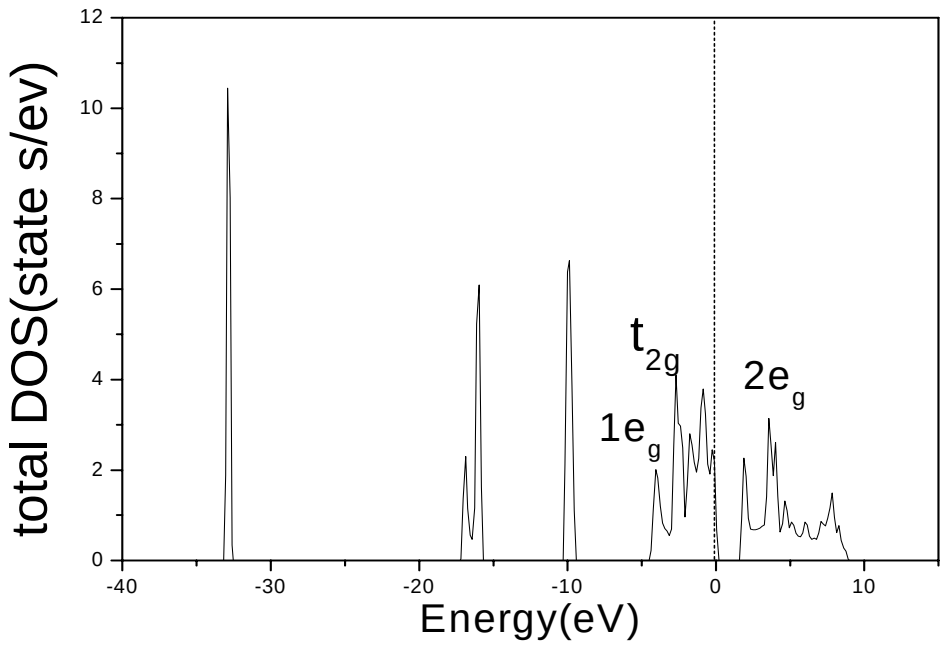

Fig.4 The total DOS Vs energy.

In order to find the composition which determined the band structure, the partial density of states (PDOS) have been employed with consideration of the spin polarization. The O 2s states positioned at -17eV and Ba 5p states at -10 eV further confirmed the band structure. Moreover, no significant differences have been detected in up and down spin states with respect to Ti atom, although slight difference may exist. From the partial DOS of O and Ti atom, it can be observed that the band near the Fermi level including the conduction band and valence band were caused by the hybridization of O 2p and Ti 3d states. Because of the octahedron centered by the Ti atom, the charges were redistributed to form the $e_g$ and $t_{2g}$ level, and they positioned up and below the Fermi level respectively. As we known, $t_{2g}$ was composed of $d_{xy}$, $d_{yz}$ and $d_{xz}$ orbitals and $e_g$ was composed of $dz^2$ and $dx^2$-$y^2$ orbitals. The Ti $e_g$ orbital has interacted with O octahedron orbitals to split into bonding orbitals and antibonding orbitals (here we use $1e_g$ and $2e_g$ to indicate the bonding and antibongding orbitals respectively in Fig. 4). However there is no interplay between the Ti $t_{2g}$ orbitals and octahedral orbitals, therefore the total DOS $t_{2g}$ state mainly attributed to the Ti $t_{2g}$ orbitals. The DOS in the vicinity of the Fermi level within the band structure of $BaTiO_3$ were attributed mostly to these octahedral hybrid orbitals. The p-d overlap can be observed by the partial DOS of O and Ti shown in Fig. 5, Fig.6 and Fig.7, and we can see the O 2p states and Ti 3d states positioned in the vicinity of the Fermi level. Meanwhile

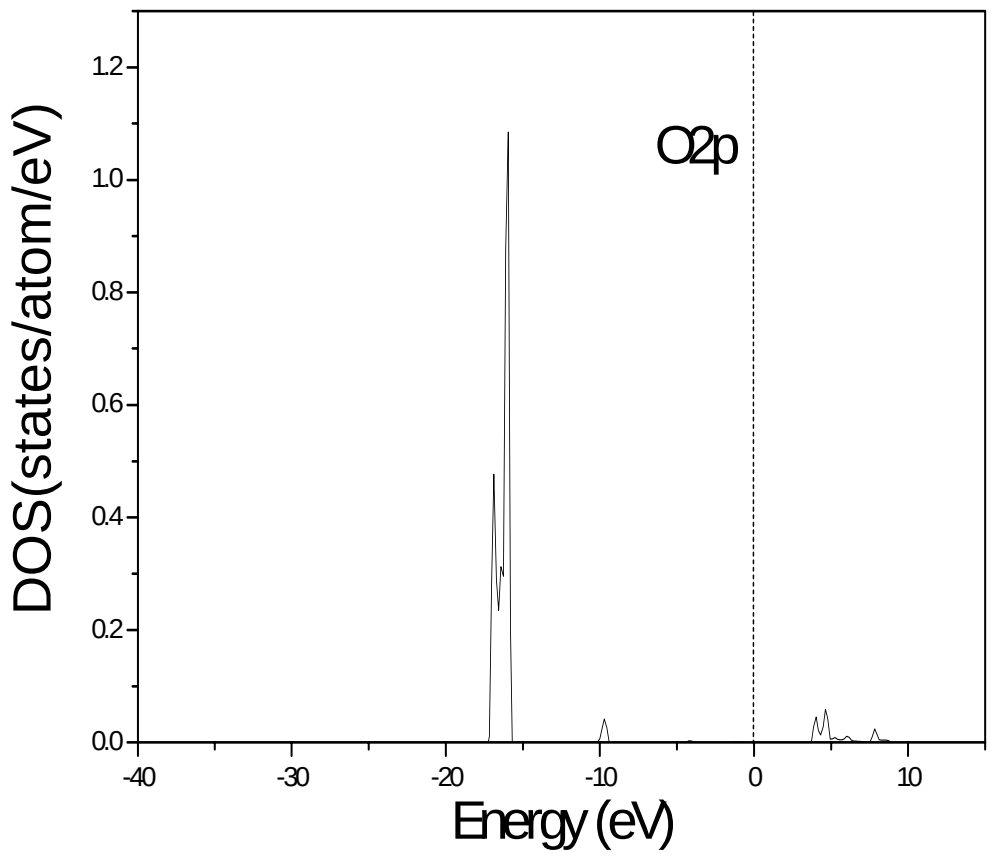

Fig. 5 Partial DOS of O Vs energy.

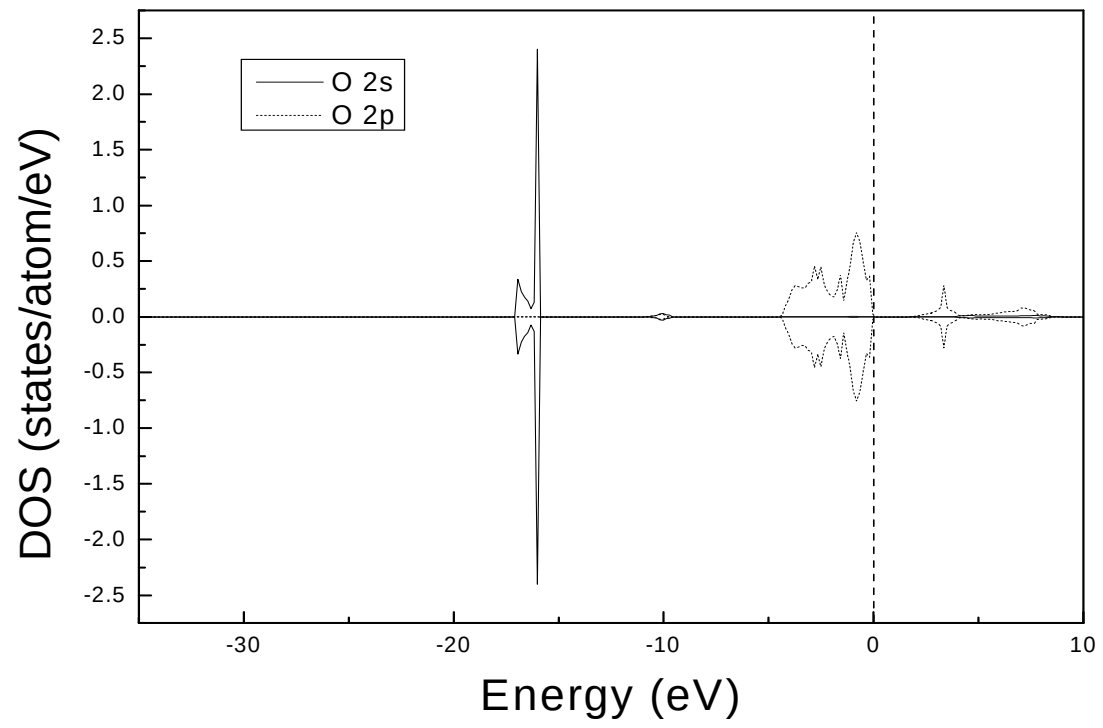

Fig.6 Spin-resolved DOS of O Vs energy

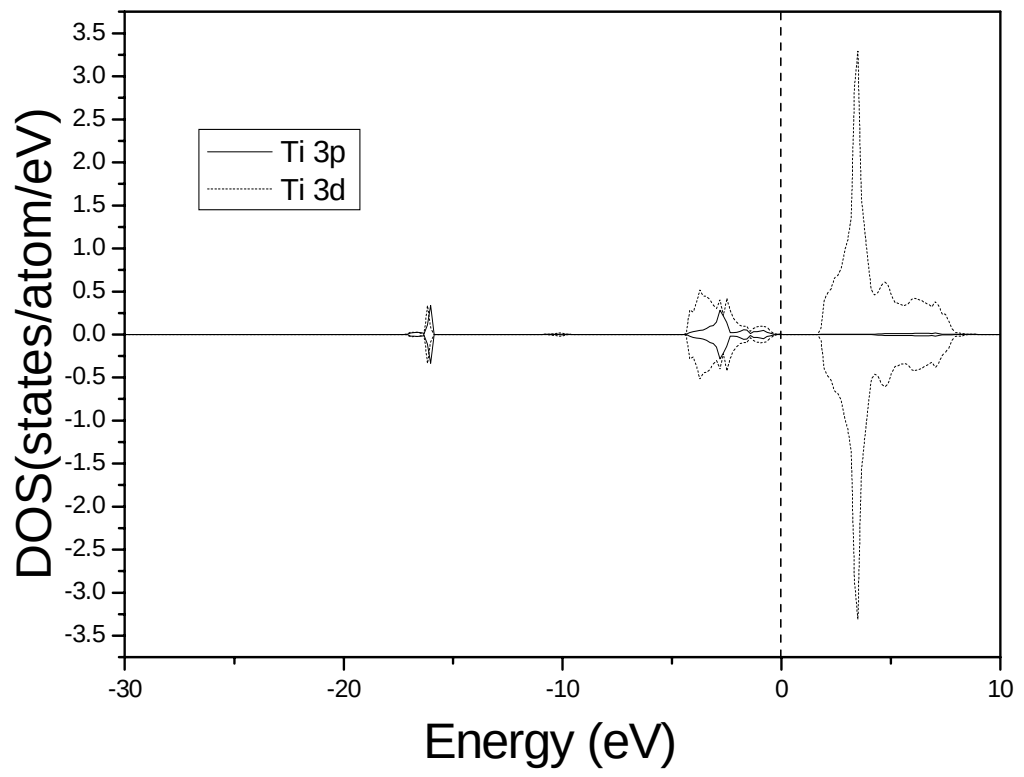

Fig. 7 Spin-resolved DOS of Ti Vs energy

the partial DOS of Ba was shown in

Fig.8, and it can be obviously seen that the band at -10 eV was mainly controlled by the Ba 5p states and the band in the topmost of the conduction

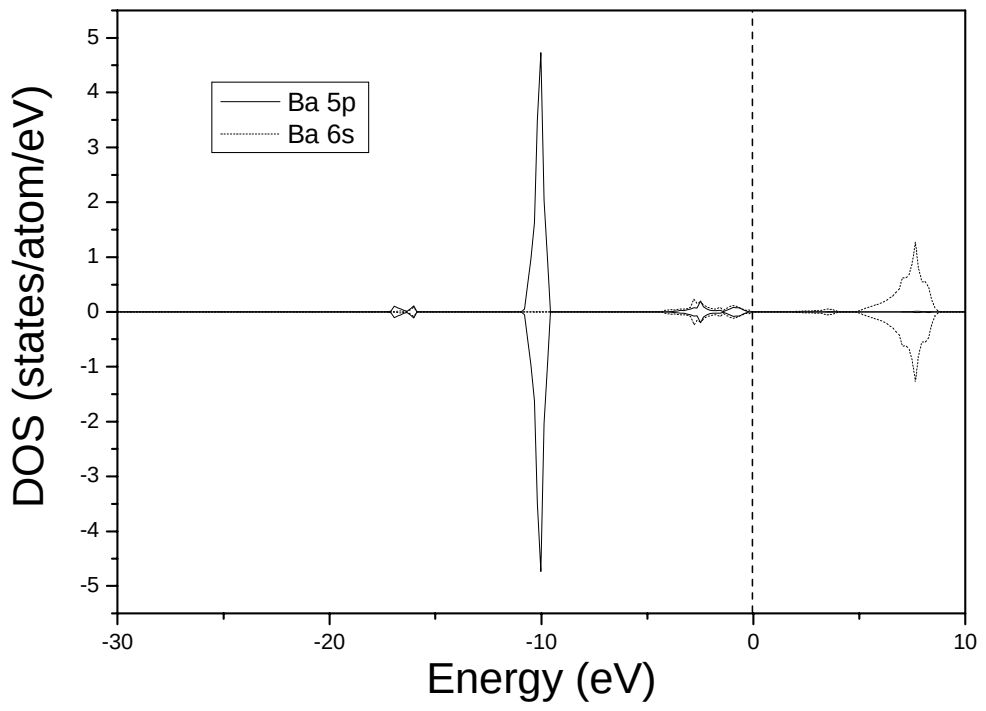

Fig. 8 Spin-resolved DOS of Ba Bs energy.

band was contributed to the Ba 6s states.

The charge density along different crystalline planes have been shown in Fig. 9, Fig.10, Fig.11 and Fig.12

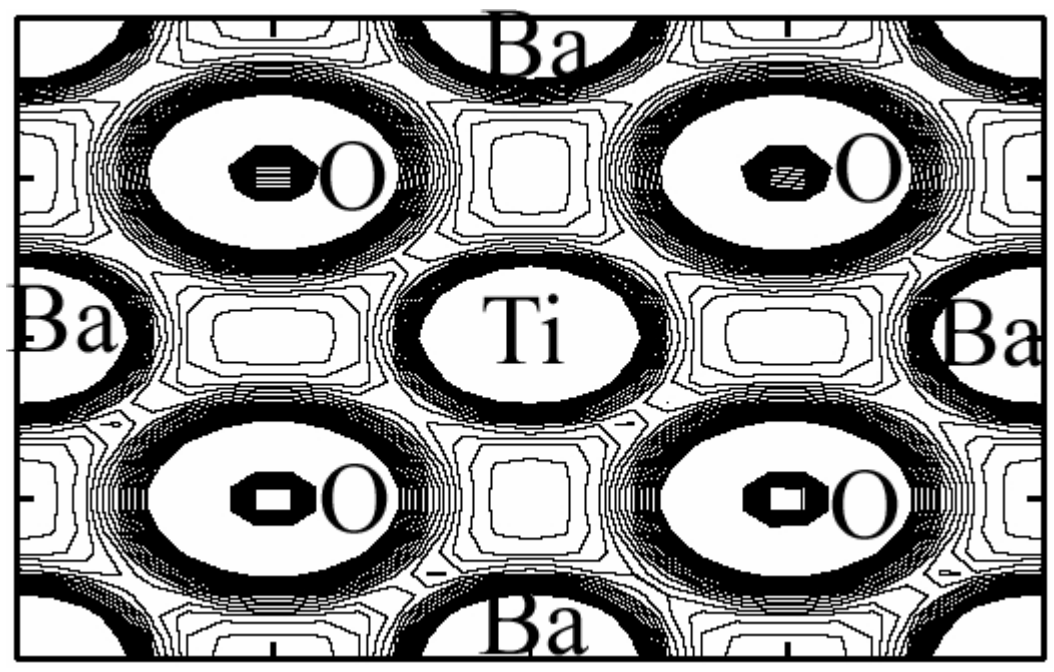

Fig. 9 The charge density distribution along (001) plane

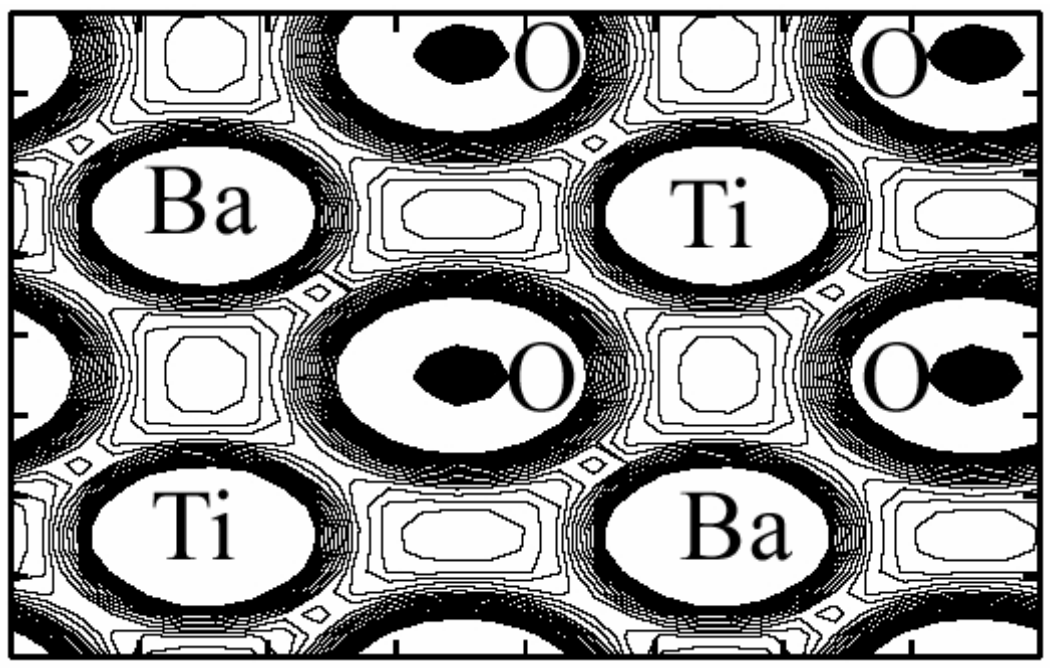

Fig. 10 The charge density distribution along (100) plane

respectively. The charge density distribution in the (001) crystalline plane and (100) crystalline plane are illustrated Fig.9 and Fig.10 respectively. There was no significant difference between these two charge density maps due to the p4mm group symmetry. In addition, the strong covalent bonding between O and Ti atom can be investigated from the maps, and this results further proved that the p-d overlap between O 2p and Ti 3d orbital. The charge redistribution along (110) crystal plane and (111) crystal plane have been shown in Fig.11 and Fig.12. The strong covalent bonding between Ti and O atoms have been affirmed again by the charge distribution along the (110) crystalline plane. Even in the (111) crystalline plane the hybridization could be observed due to the strong p-d overlap. Although most of papers published previous have made great endeavour to investigate the charge distribution of perovskites compounds, they had not projected the charge density onto different crystalline planes. Our investigation aims to show a new approach to systematically study the electronic distribution with respect to the space group symmetry.

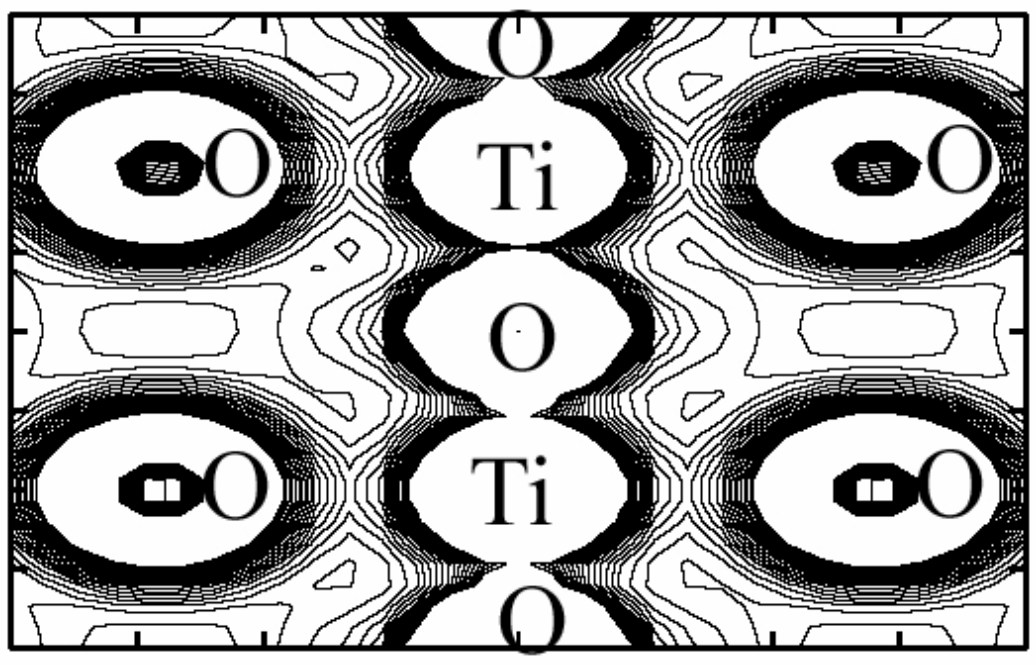

Fig.11 The charge density distribution along (110) plane

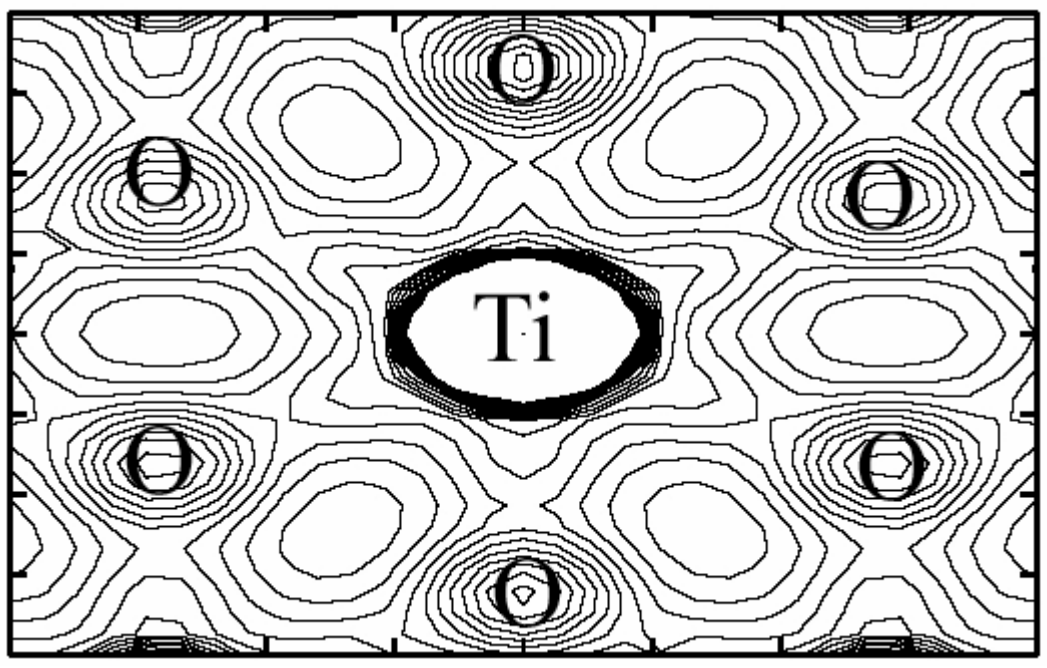

Fig. 12 The charge density distribution along (111) plane

## 4. Conclusion

Generalized gradient approximation (GGA) exchange and correlation functional as suggested by Perdew and Wang (PWGGA) within DFT has been used to study the ground states electronic properties of barium titanate. The band structure, total DOS and partial DOS have been described using the data calculated from the computation. The O 2s states positioned at -17eV and Ba 5p states at -10 eV have confirmed the lower lying quasi-flat band within the band structure. The strong interaction between O 2p and Ti 3d orbital can be confirmed from the orbital-resolved DOS of Ti and O. Moreover the Ti $e_g$ states has interplayed with the oxygen octahedral orbitals to split into bonding and antibonding orbital. The topmost band structure was mostly attributed to the Ba 6s states which can be obtained in the partial DOS of Ba. Electronic maps along (001) and (100) planes proved the covalent bonds between Ti and O, and the Ba-O bonds was found to be metallic type. Even the charge distribution projected on the (110) and (111) crystalline plane can find the bonding between Ti and O, and this point further confirm the fact that the Ti 3d orbitals interacted strongly with the octahedral oxygen to split into bonding and antibonding states.

**Acknowledgements:**

This work was supported by the Aeronautical Science foundation of China (2003ZG51069).

E.A. Kotomin, G. Borstel. Ceramics International 30 (2004) 1989-1990